\documentclass[preprintnumbers,amsmath,amssymb,aps,prl]{revtex4}

\usepackage{graphicx}
\usepackage{dcolumn}
\usepackage{bm}

\begin{document}

\preprint{}

\title{Diffusion and conduction in a salt-free colloidal suspension via molecular dynamics simulations}

\author{Sorin Bastea}
\email{sbastea@llnl.gov}
\affiliation{Lawrence Livermore National Laboratory, 7000 East Ave., Livermore, CA 94550}


\begin{abstract}
Molecular dynamics (MD) simulations are used to determine the diffusion coefficients, 
electrophoretic mobilities and electrical conductivity of a charged colloidal suspension 
in the salt-free regime as a function of the colloid charge. 
The behavior of the colloidal particles' diffusion constant can be well understood 
in terms of two coupled effects: counterion 'condensation' and slowdown due to the 
relaxation effect. We find that the conductivity exhibits a maximum which 
approximately separates the regimes of counterion-dominated and colloid-dominated conduction. 
We analyze the electrophoretic mobilities and the conductivity in terms of commonly 
employed assumptions about the role of ``free'' and ``condensed'' 
counterions, and discuss different interpretations of this approach.
\end{abstract}


\maketitle

Charged colloidal suspensions exhibit a wide range of interesting equilibrium as 
well as electrokinetic behaviors. On the equilibrium side these include charge 
renormalization due to strong counterion screening \cite{ac84,lb98,ttr08}, 
large Coulombic effects in sedimentation profiles \cite{rp04}, highly tunable 
phase transitions \cite{zr06,gns08}, solvation charge asymmetry effects \cite{kgo10}, 
{\it et caetera}. The dynamic behavior may 
be even richer due to the coupling of hydrodynamic and electrostatic 
interactions \cite{rw00,jy04,gbu08,aw08}.
The motion of charged, mesoscopically-sized particles such as colloids or polyelectrolytes 
in applied electric fields, generally referred to as electrophoresis, is relevant 
to numerous applications, from the detection of elementary charges \cite{sb08} 
to molecular biology \cite{jv00} and nanofluidics \cite{shr08}. 
As a result, the interest in elucidating fundamental aspects of this phenomenon 
remains high \cite{scc97,ew97}, with computer simulations 
playing a significant role \cite{ldh04,kn06,at08}. 
Although the focus of such studies is often only the electrophoretic mobility \cite{scc97,kn06}, 
the behavior of the self-diffusion coefficient 
and electrical conductivity are also important features of these 
systems \cite{de98,ws02}.
Electrophoretic mobility and electrical conductivity measurements are in fact routine 
characterization tools for the charge and number density of colloidal particles \cite{rp04,ws02}, 
while the determination of the diffusion constant is commonly used to obtain information 
on their size \cite{sbn06}. 
Since these are the major parameters determining the stability, structure 
and phase behavior of colloidal suspensions \cite{zr06,hh00}, unraveling the 
mechanisms controlling charge and diffusive transport and their exact relation  
to colloidal charge, size and density is crucial for understanding the properties 
of charged suspensions and how to potentially control them. Numerical simulations are 
an important avenue for elucidating such connections and new methods 
based on coupling the equations of motion of the colloidal particles 
with hydrodynamics equations for the solvent have recently proved 
valuable for studying the electrophoretic mobility \cite{kn06,at08,ch05}, while Brownian 
dynamics simulations have yielded all transport coefficients \cite{djd09}. 
Molecular dynamics (MD) simulations are a complementary tool for such studies and can 
directly capture the complex interplay of short and long range interactions as well as solvent 
structure and hydrodynamics which are relevant for transport processes in colloids, 
but have been previously limited to single colloidal particle systems \cite{ldh04,ld07,dls08}.
In this letter we present MD simulation results for the electrical conductivity, 
self-diffusion coefficients and 
electrical mobilities of a charged colloidal suspension under no-salt conditions. 
Such suspensions can provide substantial insight for example into 
surface charge regulation effects \cite{kf05}, and they have recently began 
to be studied more systematically, with results now available for both 
structural \cite{rc08} and electrokinetic properties \cite{ld07}.  

The system considered here contains solvent particles, colloidal particles with charge $-Ze$ 
and counterions of opposite unit charge, i.e. it is a salt-less suspension.
The inter-particle potentials consist of short range and Coulomb contributions.
The short range interactions are based on the inverse-12, 'soft-sphere' potential,
\begin{equation}
u(r)=\epsilon\left(\frac{d_0}{r}\right)^{12} 
\end{equation}
, which we truncate and shift at $r/d_0=2$. (We also define $u(r)=\infty$ for $r<0$.) 
They are:
\begin{subequations}
\begin{eqnarray}
&&u_{CC}(r) = u(r - 2R_C)\\
&&u_{Cc}(r) = u_{Cs}(r) = u(r - R_C)\\
&&u_{cc}(r) = u_{ss}(r) = u_{cs}(r) = u(r)
\end{eqnarray}
\label{eq:ur}
\end{subequations}
, where $C$, $c$ and $s$ denote the colloidal particles, counterions and solvent particles, 
respectively; $R_C$ is an impenetrable 
colloidal particle core radius. Such potentials have been employed 
before to model neutral suspensions \cite{sb06,sb07}.
For temperatures $k_B T\simeq \epsilon$ the effective diameters 
corresponding to these interactions are well approximated by $d_{\gamma}=d_{cs}=d_0$, 
$d_C=2R_C + d_0$, and $d_{C\gamma}=R_C+d_0$, and satisfy 
additivity, $d_{C\gamma} = (d_C+d_{\gamma})/2$; $\gamma=c, s$.
The Coulomb interactions are given by
\begin{equation}
v_{\alpha\beta}(r) = \frac{1}{4\pi\epsilon_0\epsilon_r}\frac{q_{\alpha}q_{\beta}}{r}
\end{equation}
where $\alpha,\beta=c, C$. Thus, while the solvent size granularity is explicitly 
accounted for at the microscopic level by the short range interactions, the (relative) dielectric 
constant $\epsilon_r$, as is usually the case \cite{hh00,lk08}, is not. 
It would be difficult to perform large scale simulations with solvent particles carrying 
explicit dipoles, and we do not expect that they would change the results reported here.
 
We focused on suspensions with colloid 'volume fraction' $\phi_C=\pi n_C{d_C}^3/6=0.1$, 
solvent plus counterions 'volume fraction' $\phi_0=\pi (n_c+n_s)d_0^3/6=0.35$, and 
with a colloid-solvent 'diameter' ratio $d_C/d_0=10$; $n_{\delta} (\delta=C, c, s)$ are the 
number densities and satisfy $n_c=Zn_C$ due to charge conservation.  We performed MD 
simulations of this system (which may be viewed as a model nanocolloidal suspension) 
in the microcanonical (NVE) ensemble for 8 different values of 
the colloidal charge $Z: 0, 10, 20, 30, 40, 50, 70, 100$ at an average temperature 
$k_B T=\epsilon$. If we introduce the Bjerrum length 
$\lambda_B=e^2/4\pi\epsilon_0\epsilon_rk_BT$ and Debye screening distance associated with 
counterions only, $\lambda_D = (4\pi\lambda_Bn_c)^{-\frac{1}{2}}$, the simulations 
correspond to $\lambda_B/d_0 = 2.32$, while $\lambda_D/d_0$ varied between $4.23$ for 
the $Z=10$ simulations and $1.34$ for the $Z=100$ ones. 
The particle masses were $m_s = m_c=m_0$ and $m_C/m_0 = 1000$. 
All simulations were carried out with $N_C=50$ and $N_s+N_c=175000$ in a box with 
periodic boundaries. The Coulomb interactions 
were handled using the Ewald summation technique with conducting boundary conditions. 
The time, electrical conductivity and electrical mobility units are 
$t_0=d_0(m_0/\epsilon)^{\frac{1}{2}}$, $\sigma_0=10^{-4}\times e^2t_0/m_0d_0^3$ 
and $\mu_0=et_0/m_0$, respectively.

After equilibration the simulations were run for $\simeq 5\times 10^5$ time steps ($\simeq 4000t_0$), 
accumulating structural as well as dynamic information necessary for the calculation of self-diffusion, 
electrical conductivity and electrophoretic mobilities. The self-diffusion coefficients $D_\delta$, $\delta=C, c, s$ 
were determined using the velocity autocorrelation relation:
\begin{equation}
D_\delta=\lim_{t \rightarrow \infty}\frac{1}{3}\int_{0}^{t}\langle \mathbf{v}_\delta(0)\cdot \mathbf{v}_\delta(\tau)\rangle d\tau
\end{equation}
with no tail corrections. The conductivity calculation was done 
by integrating the charge current autocorrelation, 
\begin{subequations}
\begin{eqnarray}
&&\sigma=\lim_{t \rightarrow \infty}\frac{1}{3Vk_BT}\int_{0}^{t}\langle \mathbf{j}^q(0)\cdot \mathbf{j}^q(\tau)\rangle d\tau\\
&&\mathbf{j}^q(t)=\sum_i^Nq_i\mathbf{v}_i(t)=e[\mathbf{j}^d_c(t)-Z\mathbf{j}^d_C(t)]\\
&&\mathbf{j}^d_\delta(t)=\sum_i^{N_\delta}\mathbf{v}_{i\delta}(t)
\end{eqnarray}
\end{subequations}
The electrophoretic mobilities of the colloidal particles and counterions can also 
be calculated in the Green-Kubo framework by integrating the charge - diffusion currents 
correlations \cite{dls08}:
\begin{equation}
\mu^E_\delta=\lim_{t \rightarrow \infty}\frac{1}{3N_\delta k_BT}\int_{0}^{t}\langle \mathbf{j}^q(0)\cdot \mathbf{j}^d_\delta(\tau)\rangle d\tau
\label{eq:emo}
\end{equation}
Thus $\sigma$ can be written in terms of the mobilities in its canonical form:
\begin{equation}
\sigma = e(n_c\mu^E_c - n_CZ\mu^E_C)
\label{eq:cond}
\end{equation}
It is worth noting that $\mu^E_C$ is the electrical mobility in suspension and not 
for an isolated particle, as often studied theoretically 
and measured under very dilute conditions. 

The colloid-counterion pair correlation functions - Fig. 1 - show strong counterion 
stratification at the surface of the colloidal particles and the formation of what is 
typically designated as the Stern layer \cite{hh00}, with a thickness of roughly 
one counterion diameter. This surface 'condensation' effect and the associated chemical 
equilibrium between 'condensed' and 'free' 
counterions leads to colloid-colloid interactions corresponding 
to a renormalized or effective colloid charge $Z_{eff}$, which 
saturates at large $Z$ \cite{ac84,lb98}. We test this scenario by defining $Z_{eff}$ in 
the natural way, as the charge contained on the 
average in a sphere centered on a colloidal particle and extending up to the first 
minimum of $g_{Cc}(r)$, i.e. inside the outer boundary of the Stern layer. Other, 
related definitions are also possible \cite{lub98}, but they yield similar outcomes. 
The result is plotted 
in Fig. 1 (inset) and shows that $Z_{eff}$ so defined exhibits the predicted 
saturation behavior in the range of 'bare' charges $Z$ covered in these simulations. 
Being defined in terms of $g_{Cc}(r)$, $Z_{eff}$ includes the effect of the 
counterions size. We note that since the system is salt-free 
and all counterions are identical subtle phenomena associated with 
size asymmetry between the anions and cations such as charge reversal 
and charge amplification \cite{ggo10} are not expected here. It would be very 
interesting to study the influence of such effects on the electrokinetic 
properties of charged colloids, since size-asymmetry is common 
in real situations. We should also remark that for such systems the definition of 
$Z_{eff}$ would likely need to be revised, perhaps by assuming that the charge layer 
extends up to the first minimum of the total number density of ions; 
this issue requires however further careful study. 

The self-diffusion coefficients of the colloidal particles and counterions - 
Fig. 2 (see also top inset of Fig. 3), are both decreasing functions of $Z$.
Their behavior can likely be understood by extending 
to charged colloidal systems arguments usually employed for the self-diffusion 
coefficients of electrolyte solutions. The classic analysis due 
to Onsager \cite{lo45} singles out the most important contribution to self-diffusion 
to be the relaxation effect, i.e. the drag exercised on a moving ion by its 
lagging, distorted charge atmosphere. This direct, phenomenological treatment 
is particularly suitable for the colloidal particles, which are primarily screened 
by the small counterions. In this case the ionic atmosphere subject to relaxation 
should reasonably be expected to consist only of the counterions beyond the tightly ``bound'' 
Stern layer, so an analysis in terms of the effective charge $Z_{eff}$ is likely 
appropriate; we outline it below following the ideas in \cite{lo45}.
The screening atmosphere of a charged colloidal particle moving with velocity $v$ will lag 
behind a distance $\zeta$ equal with the one traveled by 
the particle in the time $\tau$ that the atmosphere (a charge shell of typical size $\lambda_D$) 
needs to equilibrate through diffusive redistribution of the screening counterions: $\zeta=v\tau$, 
where $\tau\propto \lambda_D^2/D_c$ and $D_c=k_BT/\xi$ (Einstein relation for counterions), 
with friction coefficient $\xi\propto \eta d_c$ 
(Stokes relation with viscosity $\eta$). This lag or distortion will result 
in a retarding force between the 
charged particle and its screening atmosphere, $F_r\propto \zeta Z_{eff}^2e^2/\lambda_D^3\epsilon_r$. 
Since the Stokes drag on the colloidal particle is $F_s\propto \eta d_C v$, the total 
drag force will be 
\begin{equation}
F_{total}=F_s+F_r\propto \eta d_C v[1+f\lambda_B(d_c/d_C)Z_{eff}^2/\lambda_D]
\end{equation}
, where $f$ is a numerical factor. This corresponds to an effective friction coefficient 
\begin{equation}
\xi\prime\propto\eta d_C[1+f(d_c/d_C)Z_{eff}^2\lambda_B/\lambda_D]
\end{equation}
, which yields, according 
to the Einstein relation, the self-diffusion coefficient of the charged colloidal particles:
\begin{equation}
D_C=\frac{D_0}{1+f\frac{d_c}{d_C}\frac{\lambda_B}{\lambda_D}Z_{eff}^2}
\label{eq:donsager}
\end{equation} 
The Onsager relation for diffusion in electrolytes with equally sized anions and cations 
(whose derivation requires additional assumptions and is similar but not identical with Eq. \ref{eq:donsager}) 
also determines the factor $f$, e.g. $f\simeq0.1$ for the charge symmetric case \cite{lo45}.
For the present highly asymmetric charged colloidal system we test the relaxation 
effect prediction by assuming $f$ to be a free 
parameter. The comparison with the MD results, shown in 
Fig. 2 (inset) with $f\simeq0.022$, indicates that the concept of renormalized charge 
is suitable for describing the self-diffusive motion of charged colloids in 
conjunction with the relaxation effect. This may provide a convenient avenue for estimating 
the self-diffusion coefficients of such colloids over a wide range of charged 
states. We note that the relative diffusion constant $D_C/D_0$ satisfies the scaling 
ansatz introduced in \cite{ld07} for colloidal suspensions in the low-salt regime. 
Its behavior however is different from the standard Einstein relation usually 
assumed for charged colloids \cite{sbn06}. 

Electrical conductivity ($\sigma$) measurements are an important 
means for characterizing the properties of charged suspensions \cite{rp04,ws02}.
The MD simulations reveal that $\sigma$ exhibits a maximum as a function of the 'bare' 
charge $Z$ - Fig. 3. Such a behavior has been previously observed for the electrophoretic 
mobility of short polyelectrolyte chains \cite{gbu08}, as well as in simulations 
of a single charged colloidal particle at low volume fraction \cite{dls08,ch07}. 
In the present simulations the behavior of the counterion and colloid 
electrophoretic mobilities - Fig. 3 (bottom - inset), indicates that the conductivity maximum 
roughly separates the regimes of counterion-dominated and colloid-dominated conduction.
Theories for electrical conduction 
in classical charged systems such as electrolytes have a long history \cite{fo55,mc70} 
but their generalization to charged colloids is difficult. Common interpretations 
of electrical conduction rely on the counterion condensation effect and the assumption 
of ``free'' and ``condensed'' counterion populations that contribute differently to charge 
transport \cite{mcr05}. We examine below to what extent this picture can be reconciled with the 
observed behavior. 

Using charge conservation the conductivity equation Eq.~\ref{eq:cond} can be 
written as 
\begin{equation}
\sigma = en_CZ(\mu^E_c + \mu^E_C)
\label{eq:sig1}
\end{equation}
, where for simplicity we denote $(-\mu^E_C)$ by 
$\mu^E_C$, a positive quantity. On the other hand the assumption of ``free'' and ``condensed'' 
counterions is usually used to write the conductivity as 
\begin{equation}
\sigma = en_CZ_{eff}^*(\mu^E_{cf} + \mu^E_C)
\label{eq:sig2}
\end{equation}
, where now only $n_CZ_{eff}^*$ counterions are assumed to be ``free'', with mobility $\mu^E_{cf}$. 
This equation holds exactly if $(Z-Z_{eff}^*)$ counterions are essentially ``attached'' 
to each colloidal particle, and therefore have the exact same mobility with it. In reality however 
the ``condensed'' and ``free'' counterions are not separated populations, and the physical, 
commonly accepted picture is one of 
chemical equilibrium between them \cite{ac84}. The present simulations are consistent with this picture, 
as they show a continuous exchange between the ``condensed'' and ``free'' counterions as previously 
defined based on $g_{Cc}(r)$.
We estimate for example that the average exchange rate (per particle) is of the order $3.0\times10^{-2}t_0^{-1}$. 
Thus, as opposed to $\mu^E_c$ and $\mu^E_C$, which are given by Eq. ~\ref{eq:emo}, 
the ``free'' ions mobility $\mu^E_{cf}$ is not a very well defined quantity. Moreover, even if some reasonable 
interpretation for $\mu^E_{cf}$ is adopted, e.g. the single particle, high dilution mobility, 
$Z_{eff}^*$ can only be understood as being defined by Eqs. ~\ref{eq:sig1} and ~\ref{eq:sig2}. Thus, there is 
no {\it a priori} reason why $Z_{eff}^*$ should be exactly identified with $Z_{eff}$, although there is 
a reasonable expectation that they should behave similarly. We also note that simple assumptions 
about the relation between $\mu^E_c$ and $\mu^E_{cf}$ such as $\mu^E_c=(Z_{eff}^*/Z)\mu^E_{cf}$ are not 
compatible with Eq.~\ref{eq:sig1} unless $\mu^E_C$ is zero, or at least $\mu^E_C \ll \mu^E_c$. 

We now consider some possible interpretations of the simulation results, using the Nernst-Einstein 
relation \cite{hm}, which connects the self-diffusion constants to the mobilities and the electrical 
conductivity. 
The simplest application of this relation suggests that the conductivity, Eq.~\ref{eq:sig1}, should be written as 
\begin{equation}
\sigma = e^2n_CZ(D_c + ZD_C)/k_BT
\label{eq:ne1}
\end{equation}
, with mobilities 
\begin{subequations}
\begin{eqnarray}
&&\mu_C^E=eZD_C/k_BT\\
&&\mu_c^E=eD_c/k_BT
\end{eqnarray}
\label{eq:mob1}
\end{subequations}
Not surprisingly this is not a good approximation as it deviates rapidly 
at low $Z$ from the simulation results, and predicts a monotonously increasing conductivity - see Fig. 3 (bottom). 
The main reason for the large discrepancy is likely the implicit neglect in the above relation of 
the correlations between particles and counterions, which is not consistent with the 
existence of a tightly ``bound''   counterions layer. Such deviations from ideal behavior have also been 
noted in Brownian dynamics simulations of charged suspensions \cite{djd09}.

The Nernst-Einstein relation can also be written starting from Eq.~\ref{eq:sig2}, as 
\begin{equation}
\sigma = e^2n_CZ_{eff}^*(D_{cf} + Z_{eff}^*D_C)/k_BT
\label{eq:ne2}
\end{equation}
, with mobilities 
\begin{subequations}
\begin{eqnarray}
&&\mu_C^E=eZ_{eff}^*D_C/k_BT\\
&&\mu_{cf}^E=eD_{cf}/k_BT
\end{eqnarray}
\label{eq:mob2}
\end{subequations}
This however requires some interpretation 
on the meaning of $D_{cf}$, the diffusion constant of ``free'' counterions. We assume here that $D_c(Z=0)$ is a 
reasonable measure for $D_{cf}$, and identify it with the diffusion constant of the solvent, 
$D_s$, since the counterions 
and solvent particles are identical in the limit $Z=0$; this quantity is independent 
of $Z$, and in fact $D_c(Z=10)$ is already essentially identical with $D_s$. This assumption 
is in line with usual conductivity modeling based on Eq.~\ref{eq:sig2} \cite{ws02,mcr05}. We then identify $Z_{eff}^*$ 
with $Z_{eff}$, and plot the results in Fig. 3. This relation underestimates the conductivity results, and 
overestimates the mobility of the colloidal particles. Moreover, the qualitative behavior of these quantities 
is different than the simulations, as they both exhibit plateaus at high $Z$. We note that the observed decrease 
in mobility at high Z is in agreement with previous simulations, for example those reported 
in \cite{ch07}.

To make further progress we adopt therefore the following ansatz: we assume that the mobilities of the counterions 
and colloids $\mu_c$ and $\mu_C$ are proportional with their respective diffusion constants $D_c$ and $D_C$, and 
that the conductivity relation reduces to the original Nerst-Einstein form Eq.~\ref{eq:ne1} 
when $Z_{eff}\to Z$. The simplest 
such form is  
\begin{equation}
\sigma = e^2n_CZ_{eff}^*(D_{c} + Z_{eff}^*D_C)/k_BT
\label{eq:nea}
\end{equation}
, with mobilities 
\begin{subequations}
\begin{eqnarray}
&&\mu_C^E=e(Z_{eff}^{*2}/Z)(D_C/k_BT)\\ 
&&\mu_c^E=e(Z_{eff}^*/Z)(D_{c}/k_BT)
\end{eqnarray}
\label{eq:mob}
\end{subequations}
The $\mu_c^E$ equation is consistent with the idea that only a fraction of the counterions, $Z_{eff}^*/Z$, 
participate on average in electrical conduction as ``free'' charges. The conductivity equation 
Eq.~\ref{eq:nea} predicts a maximum, but its value is smaller than that observed in the simulations if 
$Z_{eff}^*$ is identified with $Z_{eff}$. To get better 
agreement we adopt a simple rescaling, $Z_{eff}^*=1.4Z_{eff}$. Incidentally this scaling factor 
is similar with the one previously determined by comparing 
effective colloidal charges from structural and conductivity measurements \cite{ws02}. However, 
while such a factor may signal a difference between static and dynamic effective charges (see also below), 
we do not regard this particular value as carrying fundamental significance. 
We plot the results in Fig. 4. The agreement is good for the 
counterions mobility and at least qualitative for the conductivity. On the other hand the $\mu_C^E$ 
relation appears to hold well only at high $Z$. Nevertheless, it is worth studying further 
the validity of Eq.~\ref{eq:mob}a in the effective 
charge saturation regime (high Z), since there it would imply $(\mu_C^E)^{-1}\propto Z$, which may 
perhaps be employed for a direct determination of the charge \cite{sb08}.

The above discussion highlights the inherent limitations of the effective charge 
representation and ``free'' and ``condensed'' counterions picture in the context of electrical 
conduction and electrophoretic mobilities. To the extent that it is appropriate to regard Eq.~\ref{eq:mob}b 
as implicitly defining a dynamic effective charge, it also supports the notion that static and dynamic 
definitions of the effective charge lead to different results \cite{ws02,ldh04}.
To gain further insight into why more counterions may dynamically 
behave as ``free'', as suggested by 
the above effective charge rescaling, we consider the pair correlation function between colloidal 
particles and counterions + solvent particles - $g_{Cc+s}(r)$ - Fig. 4, 
which exhibits strong layering at short distances and on intermediate 
length scales converges to $1/(1-\phi_C)$, corresponding to a higher apparent interstitial fluid 
density. We define the position ($r_b$) right after the first peak 
of $g_{Cc+s}(r)$ where this density is reached as a boundary 
fluid layer, and calculate the effective charge contained in this shell. 
We also determine the charge shell boundary $r_{\sigma}$ corresponding to $Z_{eff}^*$, 
and find that it agrees well with $r_{b}$, 
particularly at the highest $Z$'s - Fig. 4 (inset). Both are approximately half a counterion diameter 
smaller than $r_0$, the first minimum of $g_{Cc}(r)$, which defines the 
Stern layer and $Z_{eff}$. This suggests that at high $Z$, $Z_{eff}^*$ may correspond to the boundary 
fluid layer at the surface of the colloidal particles, which is 
thinner than the Stern layer. Of course such a connection is at this point somewhat 
speculative and would certainly benefit from additional study, for example by analyzing at the 
MD level the velocity correlations between the colloidal particles and the ``condensed'' counterions 
under an applied electric field \cite{ldh04,ch07}.

In sum, MD simulations of a charged colloidal suspension in 
the salt-free regime yield the diffusion constants, mobilities and conductivity, and help 
clarify the interplay between charge saturation and slowdown 
due to the relaxation effect in the diffusive motion of colloidal particles. 
The system exhibits an electrical conductivity maximum as 
a function of the colloid charge, which roughly separates the regimes of counterion-dominated 
and colloid-dominated conduction. We analyze the electrophoretic mobilities and conductivity 
using commonly employed assumptions 
about the role of the effective charge, ``free'' and ``condensed'' conterions, and discuss 
different interpretations and some of the 
limitations of this approach. We also find, in agreement with previous observations, that 
in the effective charge saturation regime the effective transported charge appears to be 
larger than the one determined by the Stern layer, and speculate that it corresponds to the boundary 
fluid layer at the surface of the colloidal 
particles. Future simulations such as the one presented here may help further elucidate this issue. 
They should also be useful for studing the dependence of the 
diffusion constants, mobilities and conductivity on the colloidal volume fraction, which has only recently 
been adressed using many particle systems \cite{kn06}. Finally, it may be interesting to study 
electrokinetic effects in charged colloidal dispersions with lower solvent dielectric constants, 
where clustering effects may play an important role \cite{sb02}. 

This work was performed under the auspices of the U. S. Department of
Energy by Lawrence Livermore National Laboratory under
Contract DE-AC52-07NA27344.

\newpage
\begin{figure}
\includegraphics{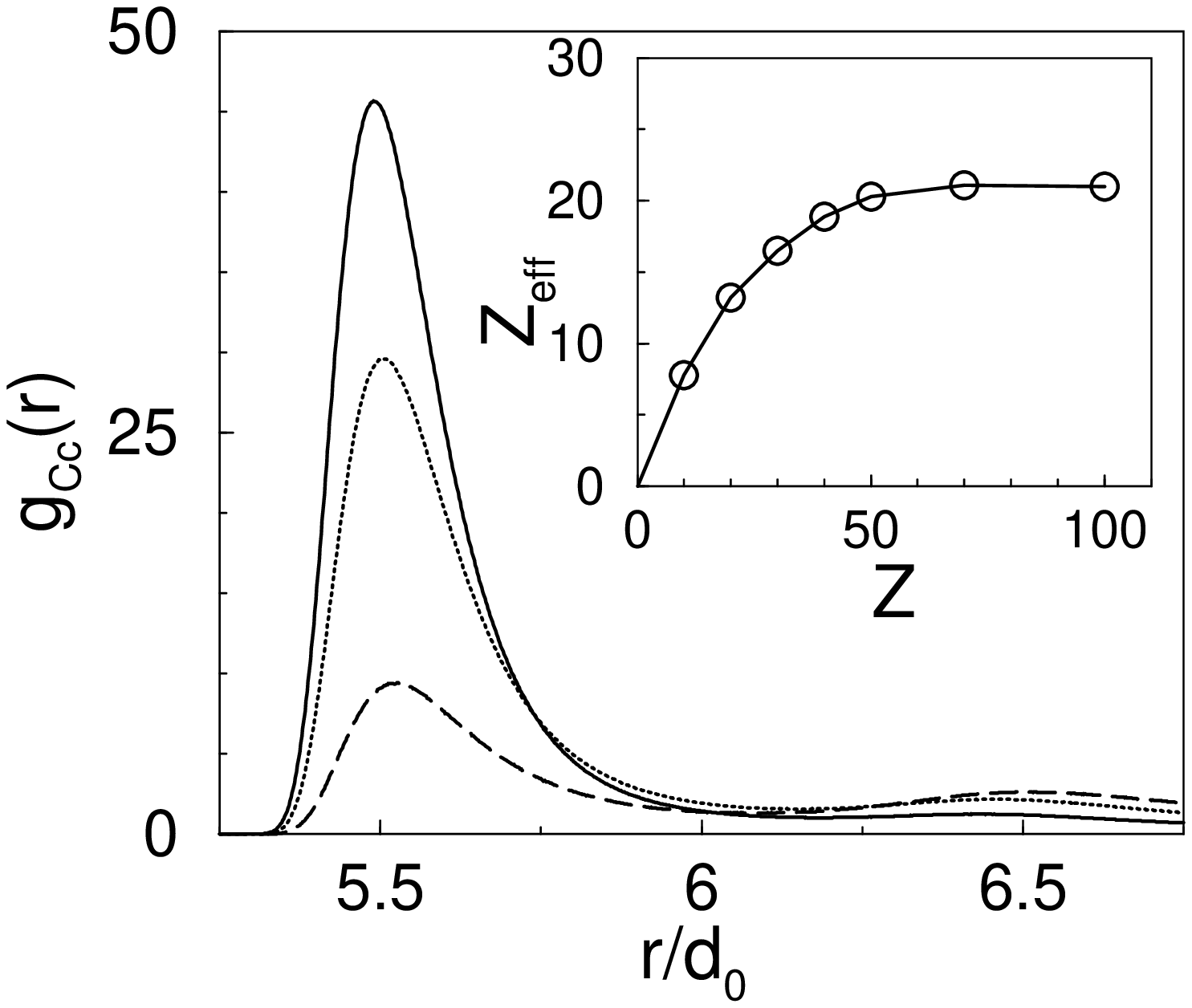}
\caption{Colloid-counterion pair correlation function for Z = 10 (dashed line), 
50 (dotted line), 100 (solid line). Inset: effective colloidal particle charge $Z_{eff}$  
as a function of the 'bare' charge $Z$.}
\label{fig:fig1}
\end{figure}

\begin{figure}
\includegraphics{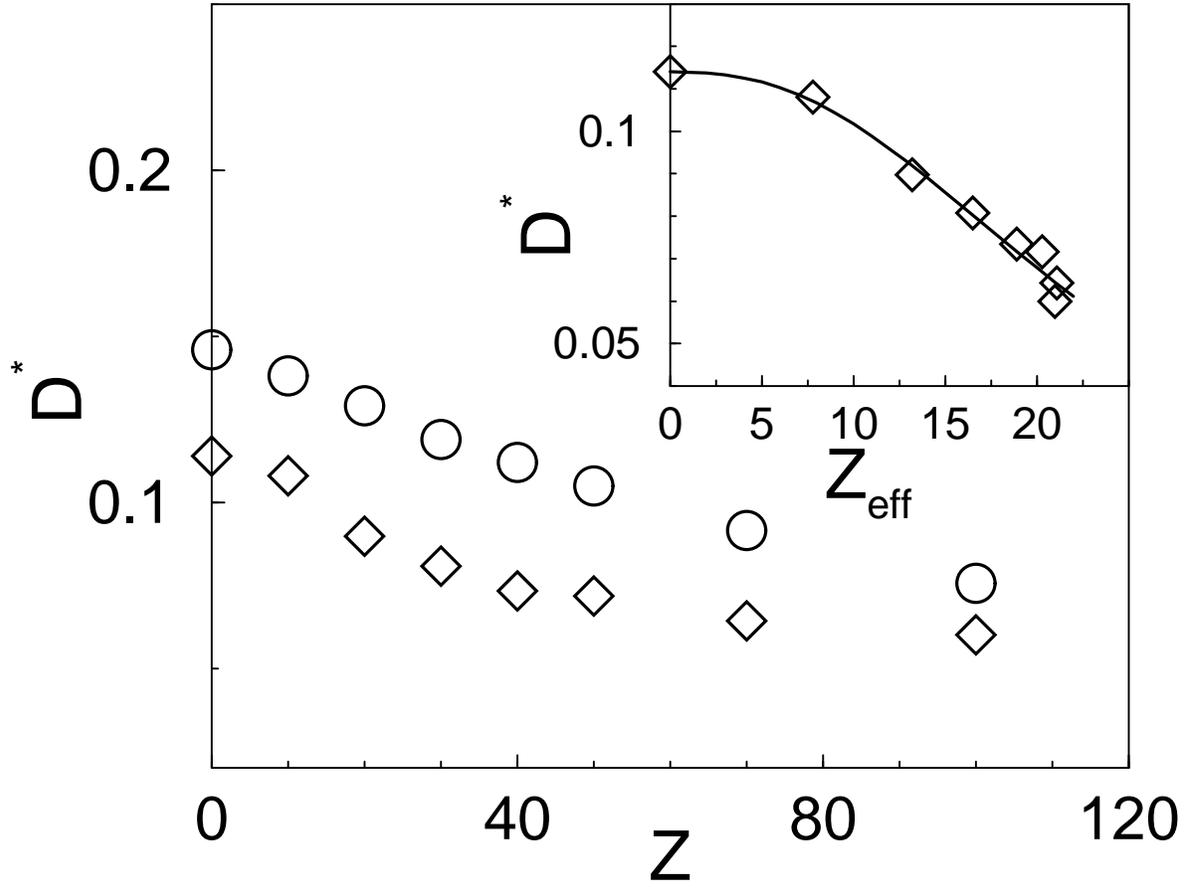}
\caption{Self-diffusion coefficients of counterions (circles) and colloidal particles 
(diamonds) as a function of colloid charge $Z$; $D^*=D_\alpha d_\alpha/D_0d_0$, $D_0=d_0^2/t_0$, 
$\alpha = c, C$. Inset: Colloidal particle self-diffusion 
coefficient as a function of the effective charge $Z_{eff}$ (symbols) and relaxation 
effect relation - Eq. \ref{eq:donsager} (solid line).}
\label{fig:fig3}
\end{figure}

\begin{figure}
\includegraphics{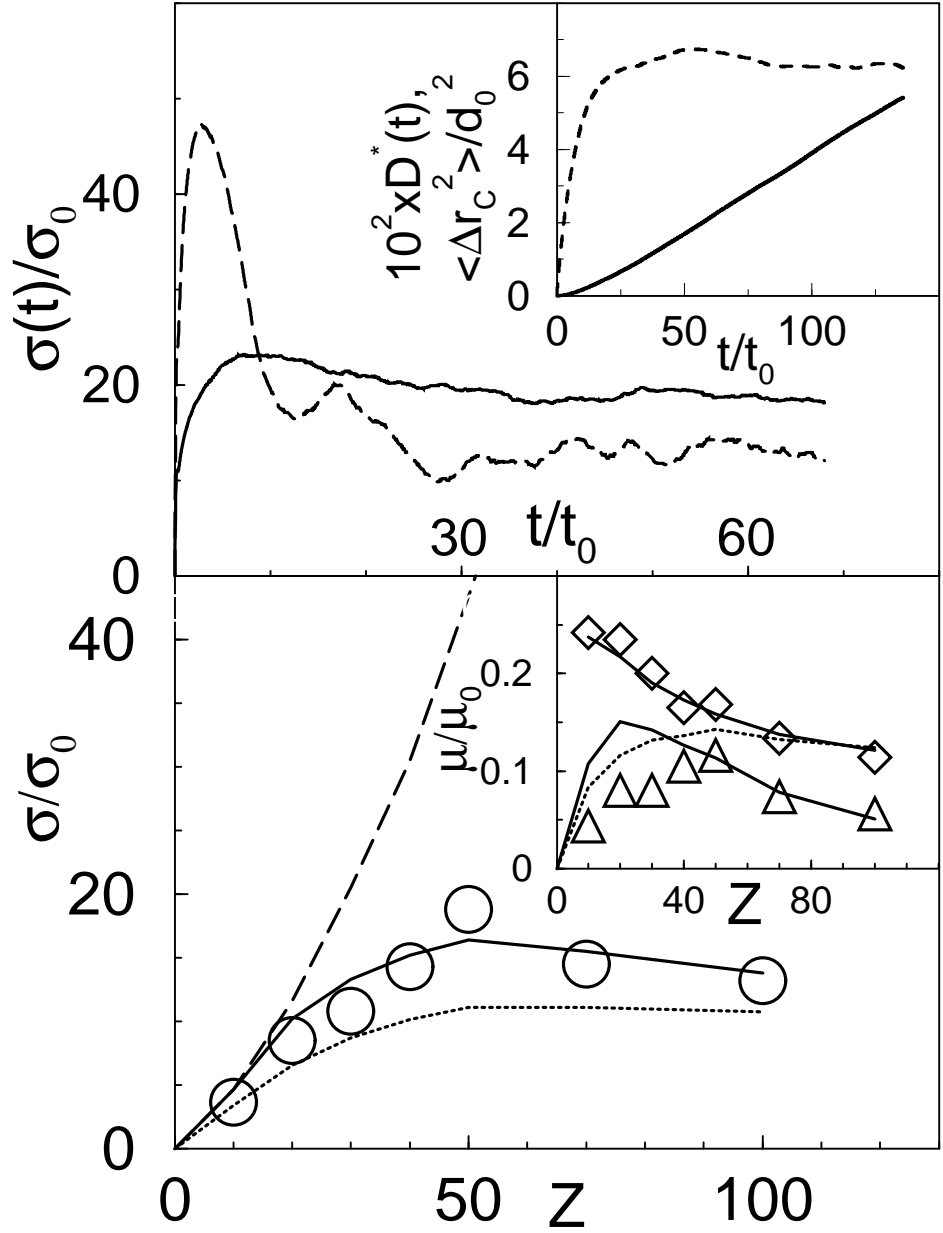}
\caption{Top: time-dependent electrical conductivity for $Z=50$ (solid line) and $Z=100$ (dashed line);
inset: time-dependent diffusion coefficient (dashed line) and mean-squared displacement (solid line) 
for colloidal particles with $Z=50$.
Bottom: electrical conductivity - simulations (circles) and 
Eq.~\ref{eq:ne1} (dashed line), Eq.~\ref{eq:ne2} with $Z_{eff}^* = Z_{eff}$ (dotted line) 
and Eq.\ref{eq:nea} with $Z_{eff}^{*}=1.4Z_{eff}$ (solid line). Inset: electrophoretic mobilities 
(absolute values) of the colloidal particles (triangles) and counterions (diamonds - shifted up 0.1 units); 
solid lines are Eqs.~\ref{eq:mob} and dotted line is $eZ_{eff}D_C/k_BT)$.}
\end{figure}

\begin{figure}
\includegraphics{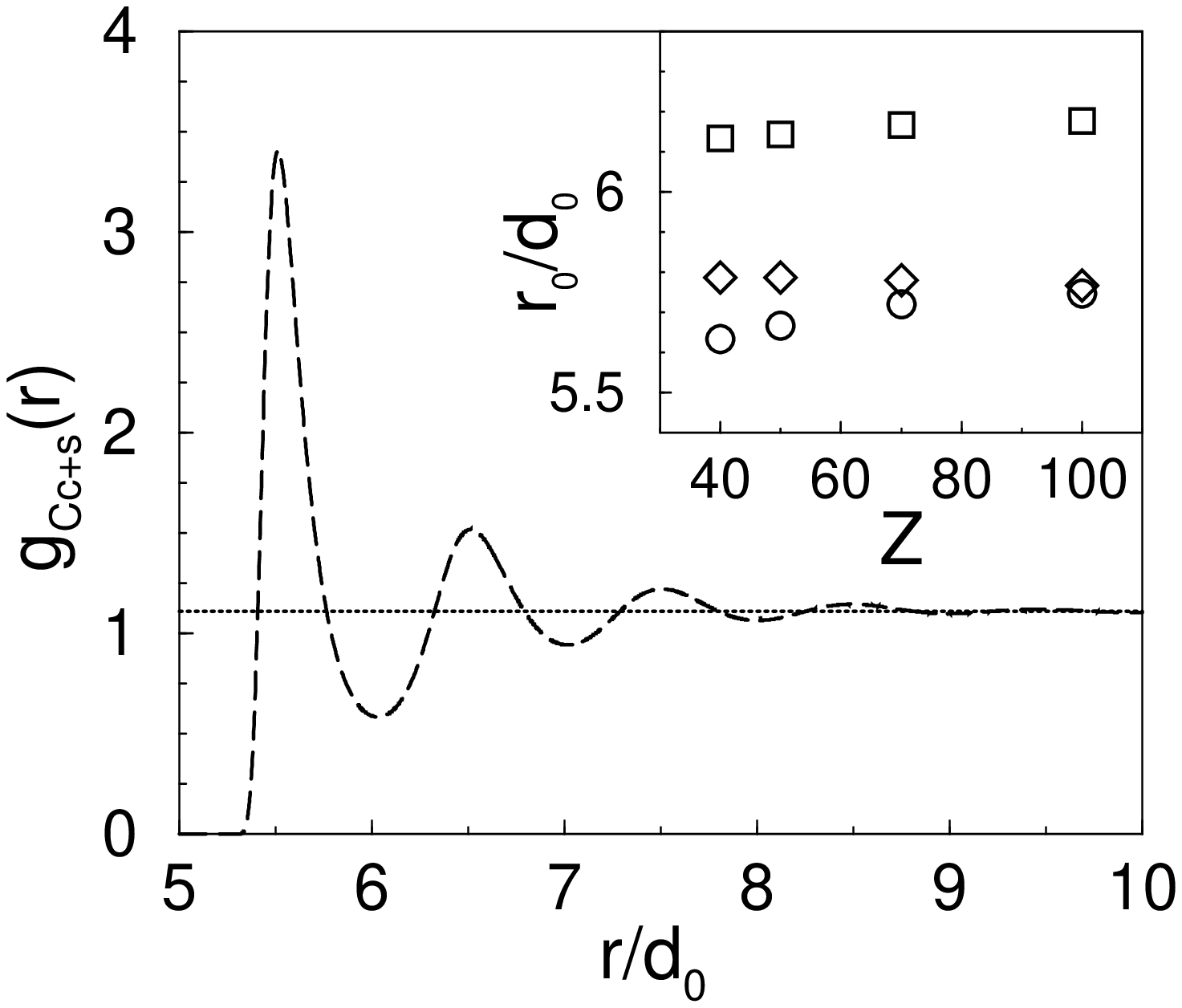}
\caption{Colloid-(solvent+counterions) pair correlation function $g_{Cc+s}(r)$ - dashed line; 
dotted line corresponds to $1/(1-\phi_C)$. Inset: first minimum of $g_{Cc}(r)$ - $r_0$ 
(squares), fluid layer boundary - $r_{b}$ (see text) (diamonds), 
and charge shell boundary corresponding to $1.4Z_{eff}$ - $r_{\sigma}$(circles).}
\label{fig:fig4}
\end{figure}


\begin{thebibliography}{99}
\bibitem{ac84}S. Alexander, P.M. Chaikin, P. Grant, G.J. Morales, P. Pincus, 
J. Chem. Phys. {\bf 80}, 5776 (1984).
\bibitem{lb98}Y. Levin, M.C. Barbosa, M.N. Tamashiro, Europhys. Lett. {\bf 41}, 
123 (1998).
\bibitem{ttr08}A. Torres, G. T\'ellez, R. van Roij, J. Chem. Phys. {\bf 128}, 
154906 (2008).
\bibitem{rp04}M. Ra\c sa, A.P. Philipse, Nature {\bf 429}, 857 (2004).
\bibitem{zr06}B. Zoetekouw, R. van Roij, Phys. Rev. Lett. {\bf 97}, 
258302 (2006).
\bibitem{gns08} H. Guo, T. Narayanan, M. Sztuchi, P. Schall, G. H. Wegdam, 
Phys. Rev. Lett. {\bf 100}, 188303 (2008).
\bibitem{kgo10}W. Kung, P. Gonz\'alez-Mozuelos, M. Olvera de la Cruz, 
Soft Matter {\bf 6}, 331 (2010).
\bibitem{rw00}D.O. Riese, G.H. Wegdam, W.L. Vos, R. Sprik, D. Fenistein, 
J.H.H. Bongaerts, G. Gr\"ubel,
Phys. Rev. Lett. {\bf 85}, 5460 (2000).
\bibitem{jy04}L. Joly, C. Ybert, E. Trizac, L. Bocquet, Phys. Rev. Lett. {\bf 93}, 
257805 (2004). 
\bibitem{gbu08}K. Grass, U. B\"ohme, U. Scheler, H. Cottet, C. Holm, 
Phys. Rev. Lett. {\bf 100}, 096104 (2008).
\bibitem{aw08}A. W\"urger, Phys. Rev. Lett. {\bf 101}, 108302 (2008).
\bibitem{sb08}F. Strubbe, F. Beunis, K. Neyts, Phys. Rev. Lett. {\bf 100}, 
218301 (2008).
\bibitem{jv00}J.-L. Viovy, Rev. Mod. Phys. {\bf 72}, 813 (2000).
\bibitem{shr08}R.B. Schoch, J. Han, P. Renaud, Rev. Mod. Phys. {\bf 80}, 839 (2008).
\bibitem{scc97}A.A. Shugai, S.L. Carnie, D.Y.C. Chan, J.L. Anderson, 
J. Colloid Interface Sci. {\bf 191}, 357 (1997).
\bibitem{ew97}J. Ennis, L.R. White, J. Colloid Interface Sci. {\bf 185}, 157 (1997).
\bibitem{ldh04}V. Lobaskin, B. D\"unweg, C. Holm, J. Phys. Condens. Matter {\bf 16}, 
S4063 (2004).
\bibitem{kn06}K. Kim, Y. Nakayama, R. Yamamoto, Phys. Rev. Lett. {\bf 96},
208302 (2006).
\bibitem{at08}T. Araki, H. Tanaka, Europhys. Lett. {\bf 82}, 18004 (2008).
\bibitem{de98}D. Erta\c s, Phys. Rev. Lett. {\bf 80}, 1548 (1998).
\bibitem{ws02}P. Wette, H.J. Sch\"ope, T. Palberg, J. Chem. Phys. {\bf 116}, 
10981 (2002).
\bibitem{sbn06}F. Strubbe, F. Beunis, K. Neyts, J. Colloid and Interface Sci. {\bf 301}, 
302 (2006).
\bibitem{hh00}J.-P. Hansen, H. L\"owen, Annu. Rev. Phys. Chem. {\bf 51}, 209 (2000).
\bibitem{ch05}A. Chatterji, J. Horbach, J. Chem. Phys. {\bf 122}, 184903 (2005).
\bibitem{djd09}V. Dahirel, M. Jardat, J.F. Dufr\^eche, P. Turq, J. Chem. Phys. 
{\bf 131}, 234105 (2009).  
\bibitem{ld07}V. Lobaskin, B. D\"unweg, M. Medebach, T. Palberg, C. Holm, 
Phys. Rev. Lett. {\bf 98}. 176105 (2007).
\bibitem{dls08}B. D\"unweg, V. Lobaskin, K. Seethalakshmy-Hariharan, C. Holm, 
J. Phys. Condens. Matter {\bf 20}, 404214 (2008).
\bibitem{kf05}R. Karnik, R. Fan, M. Yue, D. Li, P. Yang, A. Majundar, Nano. Lett. 
{\bf 5}, 943 (2005).
\bibitem{rc08}L.F. Rojas-Ochoa, R. Castaneda-Priego, V. Lobaskin, A. Stradner, 
F. Scheffold, P. Schurtenberger, Phys. Rev. Lett. {\bf 100}, 178304 (2008).
\bibitem{sb06}S. Bastea, Phys. Rev. Lett. {\bf 96}, 028305 (2006).
\bibitem{sb07}S. Bastea, Phys. Rev. E {\bf 75}, 031201 (2007).
\bibitem{lk08}T.S. Lo, B. Khusid, J. Koplik, Phys. Rev. Lett. {\bf 100}, 
128301 (2008).
\bibitem{lub98}L. Belloni, Colloids Surf. A {\bf 140}, 227 (1998).
\bibitem{ggo10}G.I. Guerrero-Garc\'ia, E. Gonz\'alez-Tovar, M. Olvera de la Cruz, 
Soft Matter {\bf 6}, 2056 (2010).
\bibitem{lo45}L. Onsager, Ann. N. Y. Acad. Sci. {\bf 46}, 241 (1945).
\bibitem{ch07}A. Chatterji, J. Horbach, J. Chem. Phys. {\bf 126}, 064907 (2007).
\bibitem{fo55}R.M. Fuoss, L. Onsager, Proc. N. A. S. {\bf 41}, 274 (1955).
\bibitem{mc70}T.J. Murphy, E.G.D. Cohen, J. Chem. Phys. {\bf 53}, 2173 (1970).
\bibitem{hm}See, e.g., J.-P. Hansen, I.R. McDonald, {\it Theory of Simple Liquids}, $2^{nd}$ edition,
(Academic Press, London, 1986).
\bibitem{mcr05}M. Medebach, R.C. Jord\'an, H. Reiber, H.-J. Sch\"ope, R. Biehl, M. Evers, 
D. Hessinger, J. Olah, T. Palberg, E. Sch\"onberger, P. Wette, J. Chem. Phys. {\bf 123}, 
104903 (2005).
\bibitem{sb02}See, e.g., S. Bastea, Phys. Rev. E {\bf 66}, 020801(R) (2002) and references therein.
\end{thebibliography}
\end{document}